\documentclass[twocolumn]{article}
\usepackage{latex8}
\usepackage{amsmath}
\usepackage{amssymb}
\usepackage{graphics}
\usepackage{alltt}
\usepackage{boxedminipage}
\usepackage{url} 

\DeclareSymbolFont{AMSb}{U}{msb}{m}{n}
\DeclareMathSymbol{\N}{\mathord}{AMSb}{"4E}
\DeclareMathSymbol{\Z}{\mathord}{AMSb}{"5A}
\DeclareMathSymbol{\R}{\mathord}{AMSb}{"52}

\begin{document}

\title{Modeling and Validating Hybrid Systems Using VDM and Mathematica}

\author{Bernhard K.\ Aichernig and Reinhold Kainhofer \\
           Institute for Software Technology (IST), \\
           Technical University Graz, 
           M\"unzgrabenstr.\ 11/II, 8010 Graz, 
           Austria}
\date{}
\maketitle
\thispagestyle{empty}
\section*{Abstract}
Hybrid systems are characterized by the hybrid evolution of their
state: A part of the state changes discretely, the other part changes
continuously over time. Typically, modern control applications belong
to this class of systems, where a digital controller interacts with a
physical environment. In this article we illustrate how a combination
of the formal method VDM and the computer algebra system Mathematica
can be used to model and simulate both aspects: the control logic and
the physics involved. A new Mathematica package emulating VDM-SL has
been developed that allows the integration of differential equation
systems into formal specifications. The SAFER example from
\cite{Kelly&97} serves to demonstrate the new simulation capabilities
Mathematica adds: After the thruster selection process, the
astronaut's actual position and velocity is calculated by numerically
solving Euler's and Newton's equations for rotation and
translation. Furthermore, interactive validation is supported by a
graphical user interface and data animation.
\section{Introduction}
Modern control applications are realized through microcontrollers
executing rather complex control logics. This complexity is increased
by the fact that control software interacts with a physical
environment through actors and sensors. Such systems are called {\it
hybrid systems} due to the hybrid evolution of their state: One part
of the state (variables) changes discretely, the other part changes
continuously over time.

Hybrid systems are excellent examples for motivating the use of
formal software development methods. First, their complexity calls for
a real software engineering discipline applying both, a process model
as well as a mathematical method. Second, these kinds of systems are
often safety-critical which justifies formal validation and
verification techniques. Third, engineers in the control domain are
educated in the use of mathematical models for designing dynamic
systems.\footnote{The same holds for software developers coming from
classical engineering disciplines.}  In our experience, the offer of a
formal method for software development is more often appreciated by
control engineers, than by software developers used to produce
short cycle products in 'Internet time'.
 
In \cite{Kelly&97} the hybrid system SAFER has been chosen by NASA in order to
introduce to formal specification and verification techniques. SAFER is an
acronym for ``Simplified Aid For EVA (Extravehicular Activity) Rescue''. It is
a small, lightweight propulsive backpack system designed to provide self-rescue
capabilities to a NASA space crewmember separated during an EVA. In
this NASA guidebook\cite{Kelly&97}, 
SAFER is specified formally in the PVS notation and 
properties are formally proved using the PVS theorem prover \cite{PVS}.
In the guidebook the dynamic aspects are used to compare the 
continuous domain model from spacecraft attitude control with the discrete 
PVS model of SAFER's control logic. It demonstrates that the two models have
the same goals: rigorous description and prediction of behavior but that the 
needed mathematics and calculation techniques are different.

In \cite{Agerholm&97,Agerholm&97b} Agerholm \& Larsen have proposed a
cheaper testing based validation approach to the SAFER example using
an executable VDM-SL model and the IFAD VDM-SL Toolbox
\cite{Jones90a,Fitzgerald&98,VDM}. They recommend the use of a specification
executor and animator for raising the confidence in a formal
model prior to formal proving. 

We agree with Agerholm \& Larsen's arguments for such a
``light-weight'' approach to formal methods in order to facilitate the
technology transfer.  Since in several industrial projects performed
at our institute a similar experience has been made
\cite{IST-TEC-99-16,IST-TEC-99-04,IST-TEC-99-20}, one of our research
areas has become the support of testing through formal methods
\cite{IST-TEC-99-01}.
 
However, neither the PVS nor the VDM-SL model of SAFER did take the
continuous physical models into account. The reason is that, in general, 
today's formal method tools are not well suited for supporting 
continuous mathematics. This paper shows a solution the problem.

In the following it is
demonstrated how an explicit discrete model can be combined with the
continuous physical model for validation and animation. With the right tool
there is no reason why a physical model should not be included in the 
validation process of a hybrid system. Just the opposite is the case: 
\cite{Agerholm&97} detected several cases where the interface to a
cut out automatic attitude hold (AAH) control unit needed further 
clarification. 
 
In this work the commercial computer algebra system Mathematica \cite{Wolfram96} has been used to overcome the gap between discrete and
continuous mathematics. A VDM-SL package has been implemented that allows to
specify in the style of the Vienna Development Method (VDM) inside Mathematica.
Thus, explicit discrete models can be tested in combination with 
differential equation systems modeling physical behavior by solving the
equations on the fly. Even 
pre- and post-condition checking is possible. Again, NASA's SAFER system
serves as the demonstrating example. The VDM-SL specification of 
\cite{Agerholm&97b} has been taken and extended with the physics involved
in SAFER, expressed through differential equations. More precisely, the 
physical behavior is movement in space,  modeled by the laws for translation 
and rotation --- Newton's and Euler's equations for three dimensional space. 
 
Beside the execution (testing) of hybrid models, Mathematica's  
front-end supports the visual validation of such systems. 
The graphical user-interface for SAFER's hand grip is implemented inside 
the computer algebra system as well as a scientific graph representing the
movement of a crew-member using SAFER. After each control cycle, 
actual physical vectors like angular velocity or acceleration can be 
inspected together with the logical status, e.g. the thrusters firing.
Finally, it is even possible to animate a sequence of performed 
control-cycles as a movie showing the SAFER representation flying.

The structure of the rest of the paper is as follows. 
First in Section \ref{safer} an overview of the SAFER system is given, which
will serve as the demonstrating example throughout the paper.
This is followed by a discussion of VDM-SL and its realization inside
Mathematica in Section \ref{vdmsl}. Then, a description of the discrete 
SAFER model is given in Section \ref{discrete}.
Section \ref{physics} explains the differential equation systems modeling 
SAFER's physics and the coordinate transformations needed.
Then, Section \ref{hybrid} introduces to the hybrid model and demonstrates
the integration of VDM-SL and differential equation systems.
Next, the validation capabilities of our approach are discussed in
Section \ref{validation} and Section \ref{enhanced}. In the final Section 
\ref{conclusion} we draw some conclusion regarding the presented work  
in particular, as well as  possible future approaches in general. 

\section{The SAFER System}
\label{safer}
The following overview of the SAFER system is based on, and partly
copied from, the NASA guidebook \cite{Kelly&97}, which describes a
cut-down version of a real SAFER system. 
\begin{figure}[t]
\begin{center}
\begin{minipage}{6.7cm}
\resizebox{6.7cm}{!}{\includegraphics{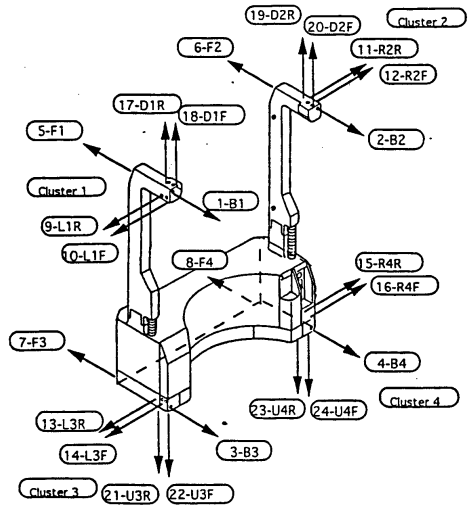}}
\end{minipage}
\caption{SAFER thrusters.}
\label{thrusters}
\end{center}
\end{figure}

The Simplified Aid for EVA Rescue (SAFER) is a small, self-contained,
backpack propulsion system enabling free-flying mobility for a NASA
crewmember engaged in extravehicular activity (EVA). It is intended for
self-rescuing on Space Shuttle missions, as well as during Space
Station construction and operation, in case a crewmember got
separated from the shuttle or station during an EVA. 
This type of contingency can arise if a
safety tether breaks, or if it is not correctly fastened. SAFER
attaches to the underside of the Extravehicular Mobility Unit (EMU)
primary life support subsystem backpack and is controlled by a single
hand controller that is attached to the EMU display and control
module. Figure \ref{thrusters} shows the backpack propulsion system
with the 24 gaseous-nitrogen ($GN_2$) thrusters, four in each of the
positive and negative $X$, $Y$ and $Z$ directions.
For example, the thrusters denoted by {\sf 5-F1, 6-F2, 7-F3} and
{\sf 8-F4} are firing backwards (indicated by the arrows) resulting in a
forward motion.  

The main focus of the discrete specification is on
the thruster selection logic, which is rather complex due to a
required priorization of hand controller commands. Various display units
and switches which are not directly related to the selection of the thrusters
have been ignored in our model. However, in contrast to \cite{Kelly&97} and
\cite{Agerholm&97} the calculation of the control output in the 
Automatic Attitude Hold (AAH) is not ignored, but simulated based on a 
dynamic model of the physics discussed in Section \ref{physics}.

\begin{figure}[h]
\begin{center}
\resizebox{6.7cm}{!}{
\includegraphics{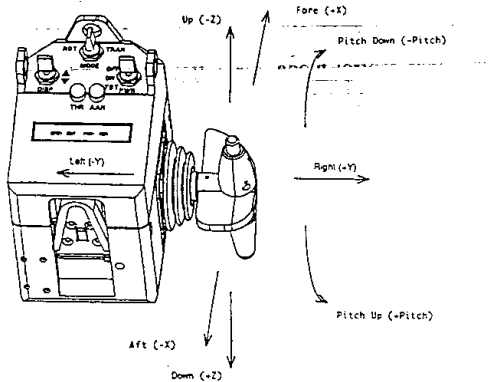}
 }
\caption{Hand controller module of SAFER.}
\label{grip}
\end{center}
\end{figure}
The hand controller, shown in Figure \ref{grip}, is a four-axis
mechanism with three rotary axes and one transverse axis using a
certain hand controller grip. A command is generated by moving the
grip from the center null position to mechanical hard-stops on the hand
controller axes. Commands are terminated by returning the grip to the
center position. The hand controller can operate in two modes, selected
via a switch, either in translation mode, where $X$ (forward-backwards),
$Y$ (left-right), $Z$ (up-down) and pitch commands are available, or in
rotation mode, where roll, pitch, yaw and $X$ commands are available.
The arrows in Figure \ref{grip} show the rotation mode commands. Note that
$X$ and pitch commands are available in both modes. Pitch commands are 
issued by twisting the hand grip around its transverse axis, while the other
commands are obtained around the rotary axis.

A push-button switch on top of the grip initiates and terminates AAH according
to a certain protocol. If the button is pushed down once the AAH is 
initiated, while the AAH is deactivated if the button is pushed twice 
within 0.5 seconds.

As mentioned above there are various priorities among commands that make the
thruster selection logic rather complicated. Translational commands issued
from the hand controller are prioritized, providing acceleration along a 
single translational axis, with the priority $X$ first, $Y$ second, and $Z$
third. When rotation and translation commands are present simultaneously from
the hand controller, rotations take higher priority and translations are
suppressed. Moreover, rotational commands from the hand grip take priority
over control output from the AAH, and the corresponding rotation axes of the
AAH remain off until the AAH is reinitialized. However, if hand grip 
rotations are present at the time when the AAH is initiated, the 
corresponding hand controller axes are subsequently ignored, until the AAH is 
deactivated.

In \cite{Agerholm&97} it is explained how a specification interpreter tool
facilitates the validation of the requirements listed in the appendix of the
NASA guidebook. Moreover, it is demonstrated that formal validation 
techniques uncover open issues in informal requirements even if they seem to
be straightforward and clear. 

The same validation techniques as discussed in \cite{Agerholm&97} can be 
applied in our Mathematica based framework --- and more. However, before we
discuss the value added through a hybrid model, in the following section, 
the realization of our VDM-SL package is discussed.
 
\section{VDM-SL in Mathematica}
\label{vdmsl}
VDM-SL is the specification language of the Vienna Development Method (VDM)
\cite{Jones90a,Fitzgerald&98}. VDM is a widely used formal method, and it 
can be applied to the construction of a large variety of systems. 
It is a model-oriented method, i.e. its formal descriptions (specifications) 
consist of an explicit model of the system being constructed. More 
precisely mathematical objects like sets, sequences and finite mappings (maps)
are used to model a system's global state. Additional logic constraints, called
data-invariants, allow one to model informal requirements by 
further restricting specified data-types. 
For validation purposes the functionality may be specified explicitly in an 
executable subset of VDM-SL. In addition, pre- and post-conditions 
state {\it what} must hold before and after the evaluation of a system's 
operation. Although VDM-SL is called a general purpose specification 
language it does not support the specification of dynamic systems. The
language's ISO-standard \cite{ISOVDM96} does not even include standard 
functions like sine or cosine.
 
Here, as the name indicates, Mathematica's strengths supplement our
combined approach. Mathematica is a symbolic algebra system that
offers the opportunity of solving arbitrary non-linear as well as
linear systems of equations. Mathematica's language interpreter is in fact a
rewriting system providing an untyped functional programming
language. For an introduction to functional programming in
Mathematica see \cite{IST-TEC-98-03}. This programming language has been 
used in order to define a package emulating the specification language VDM-SL.
By emulating we express the fact that the package does not allow one to write
specifications
in VDM-SL's concrete syntax, but in its abstract syntax with some pretty
printing for VDM-SL output. 

Mathematica's user interface are so called notebooks, fancy editors 
structured in cells for input, output or plain text. Entering a Mathematica
expression in an input cell, the system tries to evaluate this input through
a rewriting procedure based on pattern matching. 

The following language constructs have been added to the standard language in 
order to import the VDM-SL model from \cite{Agerholm&97b}:
\begin{itemize}
\item abstract datatypes for composite types, sets, sequences and maps
\item comprehension expressions for sets, sequences and maps
\item let and cases expressions
\item operators for propositional and predicate logic
\item types optionally restricted by data-invariants
\item value and global state definitions
\item typed function/operation definitions with pre- and post-conditions
\end{itemize} 
Some of the items above deserve a more detailed discussion.
\subsection*{Comprehensions}
A powerful feature of a specification language like VDM-SL is its ability to
construct
collection types like sets, sequences and maps through comprehensions.
For example, a set-comprehension defines a set through an arbitrary 
expression describing the set-elements with its free variables ranging over
a set of values, such that an optional condition holds. The following
example demonstrates the value added through a computer algebra 
system. The set-comprehension 

{\scriptsize
\begin{multline*}
\mathtt{
set[x | \{x \in \Z \} \cdot \{x^6 - 44 x^5 + 318 x^4 + 4102 x^3} \\
\mathtt{            - 4461 x^2 + 550 x + 8750 == 0\}]
}
\end{multline*}
}
represents a set of elements x, where x is an integer number such that
the equation holds. 

The resulting set\footnote{The six solutions including double and complex solutions are: $-7, -1, 1-\mbox{I}, 1+\mbox{I}, 25, 25$.}  

{\scriptsize
\[\tt set[-7, -1, 25] \]
}
demonstrates that, 
unlike IFAD's VDM-SL interpreter, comprehensions ranging over infinite
sets may be evaluated.

\subsection*{Types}
As already mentioned, in contrast to VDM, Mathematica has an untyped
language.  Consequently, no type checking mechanism is
available. However, types are an important tool for specifying a
data-model in VDM. Therefore, type declarations of the form $\tt
Type[name, type]$ have been included, where {\tt type} is one of the
predefined VDM-SL types, like basic types, composite types, sets ...
For example, a type ISet representing a set of natural numbers might
be declared by $\tt Type[ISet, set[\N]]$.

Optionally, a type can be further constrained by a data-invariant
condition. Such invariant types are defined by $\tt Type[name, type,
Invariant -> predicate]$. The {\tt predicate} is defined by a 
lambda expression mapping {\tt type} to a Boolean value.
All the invariants are globally stored in the system for invariant checking,
before and after the evaluation of a VDM function.

Internally, a type is translated to a Mathematica pattern, matching those
values the type denotes. Invariant types are supported by the possibility of
defining patterns with arbitrary predicates.
These patterns restrict the argument range in the definition of typed
VDM functions.

\subsection*{Functions}
Using the VDM-SL package, typed functions with pre- and post-conditions
can be defined using the constructor 

{\scriptsize 
\begin{verbatim}
VDMFunction[id, sig, id[vars] := body, pre, post]
\end{verbatim}
}
with the following parameters:
\begin{description}
\item[\tt id] the name of the function,
\item[\tt sig] the signature of the function,
\item[\tt \mbox{id[vars] := body}] the function definition,
\item[\tt pre] an optional pre-condition stating what must hold before the
evaluation such that the post-condition holds,
\item[\tt post] an optional post-condition stating what must hold after the
evaluation.
\end{description}
VDMFunction realizes a complex call to Mathematica's internal Function call and
emulates the checks for
\begin{itemize}
\item the signature types,
\item pre- and post-condition,
\item data-invariants.
\end{itemize} 
\section{Discrete Model}
\label{discrete}
In order to demonstrate the Mathematica package the same functions for the 
thruster selection logic as in \cite{Agerholm&97} are presented in this
section.
The six degree-of-freedom of the translation and rotation commands is modeled
using a composite type:

{\scriptsize
\begin{alltt}
Type[SixDofCommand, Composite[\{"tran", TranCommand\},
                              \{"rot",  RotCommand \}]]
\end{alltt}
}
whose two fields are finite maps from translation and rotation axis 
respectively to axis commands. For example the type of translation commands
is defined as follows:

{\scriptsize
\begin{verbatim}
Type[TranCommand, TranAxis -> Axiscommand,
     Invariant -> (dom[#] == set[X,Y,Z]&)]
\end{verbatim}
}
where the invariant ensures that command maps are total. Here, the 
invariant predicate is defined by a lambda expression in Mathematica's notation
of pure functions.
The type of 
rotation commands is defined similarly. Enumerated types are used for axis
commands and translation and rotation axes:

{\scriptsize
\begin{verbatim}
Type[AxisCommand, NEG | ZERO | POS];

Type[TranAxis, X | Y | Z];

Type[RotAxis, ROLL | PITCH | YAW]
\end{verbatim}
}

In the {\tt SelectedThrusters} function in Figure \ref{selected} grip
commands from the hand controller (with six-degree-of freedom
commands) are integrated with the AAH control output. The {\tt
IntegratedCommands} function prioritizes hand controller and AAH
commands.

\begin{figure}[t]
\begin{center}
\begin{boxedminipage}[h]{3.0938in}
\scriptsize
\begin{alltt}
VDMFunction[
  SelectedThrusters,
    AUX`SixDofCommand \(\times\) AUX`RotCommand \(\times\) 
    set[AUX`RotAxis] \(\times\) set[AUX`RotAxis] 
      -> ThrusterSet, 
  SelectedThrusters[hcm, aah, actAxes, ignHcm] :=
  let[\{tran, rot, bfMandatory,bfOptional,
     lrudMandatory,lrudOptional,bfThr,lrudThr\},
    \{tran, rot\} = 
            (IntegratedCommands[hcm,aah,actAxes,ignHcm] 
                   /. SixDofCommand[tr_,ro_]:->{tr,ro});
    \{bfMandatory, bfOptional\} = BFThrusters[tran[X],
                                            rot[PITCH], 
                                            rot[YAW]]; 
    \{lrudMandatory, lrudOptional\} = 
                                 LRUDThrusters[tran[Y],
                                               tran[Z],
                                               rot[ROLL]];
    bfThr = If[(rot[ROLL] === ZERO),
                bfOptional \(\cup\) bfMandatory,
                bfMandatory ];
    lrudThr = If[(rot[PITCH] === ZERO) and 
                 (rot[YAW] === ZERO),
                  lrudOptional \(\cup\) lrudMandatory,
                  lrudMandatory];
  set @@ (bfThr \(\cup\) lrudThr) 
  ]  
];
\end{alltt}
\end{boxedminipage}
\end{center}
\caption{The SelectedThrusters function.}
\label{selected}
\end{figure}

\begin{figure}[t]
{
\scriptsize
\begin{center}
\begin{boxedminipage}[h]{3.0938in}
\begin{alltt}
VDMFunction[
  BFThrusters,
    AUX`AxisCommand \(\times\) AUX`AxisCommand \(\times\) AUX`AxisCommand  
      -> ThrusterSet \(\times\) ThrusterSet, 
  BFThrusters[A, B, C] :=
    cases[\{A, B, C\},
      \{NEG, ZERO, ZERO\} -> \{\{B4\}, \{B2,B3\}\},
      \{ZERO, ZERO, ZERO\} -> \{\{\}, \{\}\},
      \{POS, NEG, ZERO\} -> \{\{F1,F2\}, \{\}\},
      ... 
    ]  
];

VDMFunction[
  LRUDThrusters,
    AUX`AxisCommand \(\times\) AUX`AxisCommand \(\times\) AUX`AxisCommand  
      -> ThrusterSet \(\times\) ThrusterSet, 
  LRUDThrusters[A, B, C] :=
    cases[\{A, B, C\},
      \{NEG, NEG, ZERO\} -> \{\{\}, \{\}\},
      \{NEG, ZERO, ZERO\} -> \{\{L1R,L3R\}, \{L1F,L3F\}\},
      \{POS, ZERO, POS\} -> \{\{R2R\}, \{R2F,R4F\}\},
      ... 
    ]  
];
\end{alltt}
\end{boxedminipage}
\end{center}
}
\caption{Extracts from BFThrusters and LRUDThrusters.}
\label{bfthrusters}
\end{figure}
Based on these  commands, thrusters for back and forward
accelerations and left, right, up and down accelerations are calculated by
two separate functions. Figure \ref{bfthrusters} presents cut-down versions 
of these functions.
These represent a kind of look-up tables, modeled
using cases expressions. Note that they return two sets of thruster names,
representing mandatory and optional settings respectively. 
\section{Physics Involved in SAFER}
\label{physics}
This section presents the continuous model of the physics involved in
our hybrid model. For the SAFER example, translation and rotation
equations from mechanics are sufficient for modeling the motion of a
crewmember using the propulsion system.  The purpose of this model is
twofold: First, we need to calculate the sensor inputs of angular
velocity for simulating the AAH. Second, in order to visualize
the SAFER movement, absolute coordinates have to be determined.  The 
mathematics needed can be found in the standard literature of
mechanics, like \cite{Hauser65}.

\subsection*{Translation}
The translation of a crewmember wearing SAFER is described by Newton's 
second law of motion expressed by
\begin{equation}
F = m \Dot{v} = \Dot{p} 
\end{equation}
where $F$, $m$, $v$ and $p$ denote force vector, mass, velocity vector and 
impulse vector. 
It states that ``The time rate of change of the momentum of a particle
is proportional to the force applied to the particle and in the
direction of the force.'' 
\subsection*{Rotation}
The rotation is modeled by three equations known as the 
{\it Euler's equations} of motion for the rotation of a rigid body.

Denote by $\Omega$ the angular velocity defined with respect to the center of
mass, and by $I$ the moments of inertia. The equations describing the body
rotations are then given by 
\begin{align}
I_1 \Dot{\Omega}_1 + (I_3 - I_2) \Omega_2 \Omega_3 & = Q_1 \\ 
I_2 \Dot{\Omega}_2 + (I_1 - I_3) \Omega_3 \Omega_1 & = Q_2 \\ 
I_3 \Dot{\Omega}_3 + (I_2 - I_1) \Omega_1 \Omega_2 & = Q_3 
\end{align}
or as a vector equation where $I$ is a diagonal matrix:
\begin{equation}
I \cdot \Dot{\Omega} + \Omega \times  I \cdot \Omega = Q
\end{equation}
$Q_i$ denotes a torque causing a rotation around 
the $i$-axis, in the body's own coordinate
system. Here, the torque is given by the sum over the thrusters firing. 
Actually, a component $Q^{th}$ is calculated by the cross product
of a thruster's position vector relative to the center of mass and its 
force. 
SAFER does not use proportional gas jets, but thrusters whose valves are open
or not, which simplifies the calculation. 

\subsection*{Motion}
In order to combine translation and rotation in a single model of
motion, suitable for our purposes, coordinate transformations are
necessary.  More precisely, the fixed coordinate system values for
visualization (position and velocity) have to be related to SAFER's
coordinate system values (angular velocity).

As $\Omega$ is calculated in the body's own coordinate system, they
have to be transformed back to the fixed coordinate system. Given the
Euler angles $\varphi$, $\theta$ and $\psi$ that denote the deviation
of the fixed $x$, $y$ and $z$ axis, the angular velocities can be calculated
according to the following formula.
\begin{align}
\Omega_1 & = \Dot{\varphi} \sin \theta \sin \psi + \Dot{\theta} \cos \psi \\ 
\Omega_2 & = \Dot{\varphi} \sin \theta \cos \psi - \Dot{\theta} \sin \psi \\ 
\Omega_3 & = \Dot{\varphi} \cos \theta + \Dot{\psi} 
\end{align}
The derivation of these equations
can be found in \cite{Hauser65}.
Using vector notation we get the equation:
\begin{equation}
\Omega = D_3(\psi) \cdot D_1(\theta) \cdot (\Dot{\theta}, 0,\Dot{\varphi} )^T 
                                          + (0, 0, \Dot{\psi})^T
\end{equation}
\begin{align}
           D_1 & = \begin{pmatrix}
                    1  & 0             & 0           \\
                    0  & \cos \theta   & \sin \theta \\
                    0  & - \sin \theta & \cos \theta
                   \end{pmatrix} \\
           D_3 & =  \begin{pmatrix}
                    \cos \psi    & \sin \psi & 0 \\
                    - \sin \psi  & \cos \psi & 0 \\
                    0            & 0         & 1   
                   \end{pmatrix}
\end{align}
where $D_1$ and $D_3$ are rotation matrices that 
turn the coordinate system by a given angle.

$D_1$ and $D_3$ are used to transform a vector from our
fixed coordinate system to a turned coordinate system. For translation motion,
the thruster's force vector $F$ has to be transformed from SAFER's 
coordinate
system to the fixed one using the transposed rotation matrices:
\[
(D_3(\psi) \cdot D_1(\theta) \cdot D_3(\varphi))^T
\]
Summarizing, these four vector differential equations are sufficient
for modeling SAFER's motion over time:
\begin{align}
& v    = \Dot{x} \\
& m \cdot v  = (D_3(\psi) \cdot D_1(\theta) \cdot D_3(\varphi))^T F \\
& I \cdot \Dot{\Omega} + \Omega \times  I \cdot \Omega  = Q \\
& \Omega  = D_3(\psi) \cdot D_1(\theta) \cdot (\Dot{\theta}, 0,\Dot{\varphi} )^T
                                          + (0, 0, \Dot{\psi})^T
\end{align}
Solving these equations with given thruster forces results in SAFER's position
vector $x(t)$ and the angular velocity $\Omega(t)$ used for AAH. 

Alternatives to the Euler's equations model are possible. For example,
an aproach could have involved the less computationally intensive quaternions.
However, for validation purposes the model should be as intuitive as possible,
here efficiency plays a minor role.

\section{A Hybrid Model}
\label{hybrid}
The hybrid model of SAFER consists of the hand controller and the Automatic 
Attitude Hold as its discrete parts on one side and the equations of motion 
as the continuous part on the other side. Both are modeled in Mathematica,
the first in the form of the VDM-SL specification using our VDM-SL emulation 
package, the later in the form of ordinary differential equations in Mathematica 
notation.

The combination of the discrete control system and the continuous physical 
model during the testing phase carries certain advantages:

Not only can the system specification be tested in an (idealized) physical
simulation, but also the system parameters like the force of the thrusters
and the moments of inertia of the backpack can easily be adjusted until the
system responds in a way suitable for practical use.

This is not a very rigorous approach, and it is not intended to replace 
other testing tools and methods. Rather it can serve as a valuable 
supplementary tool.

\subsection*{The Control Cycle}
The {\tt ControlCycle} function (Figure \ref{controlcycle}) integrates
 the discrete model of hand control, thruster selection and Automatic
Attitude Hold (AAH) with the continuous physical model of motion presented 
above.

The Control Cycle is implemented in two different functions. {\tt
ControlCycle} takes the state of the hand control (switches and hand
grip) as well as the already calculated or manually entered AAH
commands and the sensor values.  {\tt SensorControlCycle} takes the
values of the sensors (here simulated by the solutions of the
equations of motion of the previous control cycle) and determines
which thrusters are invoked by the AAH. These are then passed on to
{\tt ControlCycle}.

\begin{figure}[t]
\begin{center}
\begin{boxedminipage}[h]{3.0938in}
\scriptsize
\begin{alltt}
VDMFunction[
  ControlCycle,
  SwitchPositions \(\times\) HandGripPosition \(\times\)
     RotCommand \(\times\) InertialRefSensors -> ThrusterSet,

  ControlCycle[SwitchPositions[mode_, aah_], rawGrip, 
               aahCmd, IRUSensors]:=
    let[\{
      gripCmd=HCM`GripCommand[rawGrip, mode],
      thrusters=SelectedThrusters[gripCmd, aahCmd,
          AAH`ActiveAxes[], AAH`IgnoreHcm[]]
    \},
    AAH`Transition[IRUSensors, aah, gripCmd, SAFER`clock];
    SAFER`clock=SAFER`clock+1;
    PosData=CalcNewPosition[thrusters];
    thrusters
  ],
  True,
  card[RESULT] \(\leq\) 4 \(\land\) ThrusterConsistency[RESULT]
];

VDMFunction[    
  SensorControlCycle,
  SwitchPositions \(\times\) HandGripPosition -> ThrusterSet,

  SensorControlCycle[SwitchPositions[mode_, aah_], 
                     rawGrip]:=   
    ControlCycle[SwitchPositions[mode,aah],rawGrip, 
      AAHControlOut[Sensors], Sensors ]
];
\end{alltt}
\end{boxedminipage}
\end{center}
\caption{The ControlCycle function.}
\label{controlcycle}
\end{figure}

After determining the active thrusters and the AAH state, the
differential equations are solved numerically in the {\tt
CalcNewPosition} function and the current position is updated.
These results simulate the values measured by the
sensors (with exception of the heat sensors, which are left out in our
model) providing data for AAH.  
This part of the control system is completely
left out in \cite{Agerholm&97} and only included in the form of two
unspecified functions in the PVS model \cite{Kelly&97}.

Here the SAFER state is not as trivial as in \cite{Agerholm&97} where
it holds only a clock variable.
{\scriptsize
\begin{alltt}
State[SAFER,
  Type[clock, \(\N\)],
  Type[PosData, PositionData],
  Type[Sensors, InertialRefSensors],
  Type[step, \(\R\){pos}],
  Type[PosDataList, List[PositionData]],

  init[SAFER] := SAFER[0,
      PositionData[0, 0, 0, 0, 0, 0, 0, 0, 0, 0, 0, 0],
      InertialRefSensors[0, 0, 0, 0, 0, 0, 0, 0, 0],
      1/4, \{\{\{0, 0, 0\}, \{0, 0, 0\}, \{0, 0, 0\}, \{0, 0, 0\}\}\}]
];
\end{alltt}
}
The state above also includes the current position, Euler
angles and velocities stored in a variable of type {\tt PositionData}.

Even the past position data is stored for providing full information
about SAFER's trajectory. For simulation this data will be used to
display the history as a Mathematica "movie" showing the astronaut
flying around in the coordinate system.

\subsection*{Automatic Attitude Hold (AAH)}
Simulating the measured sensor values by the results of the equations
of motion provides the opportunity of including the Automatic Attitude
Hold mechanism by a simple Bang Bang \cite{Kelly&97} algorithm: If the
angular velocity for an axis where AAH is turned on exceeds a certain
threshold, selected thrusters are fired in order to slow down this rotation
(Figure \ref{bangbang}). AAH is limited to this mechanism because
SAFER is only based on simple thrusters with two states: on and off.

\begin{figure}[t]
\begin{center}
\begin{boxedminipage}[h]{3.0938in}
{\scriptsize
\begin{alltt}
VDMFunction[
  AAHControlOut,
  InertialRefSensors->RotCommand,
        
  AAHControlOut[IRUSensors]:=
    let[\{rr=IRUSensors."RollRate", 
          pr=IRUSensors."PitchRate",
          yr=IRUSensors."YawRate"\},
      map[
        ROLL->Which[
            rr \(\leq\) -EpsRoll,POS,
            rr \(\geq\) EpsRoll, NEG,
            True, ZERO],
        ...
      ]]
];
\end{alltt}
}
\end{boxedminipage}
\end{center}
\caption{The Bang Bang algorithm for AAH.}
\label{bangbang}
\end{figure}

\subsection*{The Differential Equations}
The equations of motion used to
determine the new position of the astronaut are Newton's and Euler's equations
described above. Although this model neglects any gravitational forces and other
disturbing influences, they could easily be added by an additional
acceleration in the equations or random fluctuations applied to the results of the differential
equations.

The new position is obtained by numerically solving the equations
rather than algebraically which is less time-efficient, beside the
fact that the algebraic solution is not necessary as only the result at time
{\tt step} is needed for simulation.
 
Since the equations are only slightly coupled, they can be solved in four
steps, which is numerically more stable than solving them all at once. 
This functionality is provided by Mathematica's {\tt NDSolve} function,
which takes the differential equations and the initial conditions and returns
numeric functions that approximate the exact solutions of the equations. 
In this case the trajectory is calculated piecewise: in every control 
cycle the trajectory only for that cycle is solved using the position 
before the cycle as the initial conditions and the force and torque applied 
by the thrusters as parameters. These can easily be calculated, the force 
by a simple vector addition of the forces applied by every single thruster, 
and the torque by adding up the cross products of the thruster positions 
with the force applied by that thruster.

First, Euler's equation in the astronaut's coordinate system is solved giving 
the angular velocity. This needs the forces and the torque applied by the 
fired thrusters as parameters.
The result is then transformed back to the fixed
coordinate system and used to solve the differential equation for the Euler
angles. In a third step Newton's equation can be solved using the results from
the previous equations. Finally, a simple integration of the velocities gives the
position of the astronaut.

These numerical solutions to the differential equations can also be used to 
investigate stability. In the simplified case without any external forces 
like gravitation, this might not be so interesting, but as soon as external 
forces are modeled into the differential equations, stability is a crucial 
concern. What happens if the hand controller keeps in the same position 
over a long period of time? 
Such questions can easily be answered by solving the differential 
equations for a time period longer than just the control cycle.

%
\section{Simulating SAFER}
\label{validation}
Mathematica does not only provide algebraic and numeric functionality, 
but also an extensive repertoire of plotting functions. Thus Mathematica 
has also been used to visualize SAFER's current position together with other 
state information.

\begin{figure}[h]
\begin{center}
\resizebox{3.31in}{!}{\includegraphics{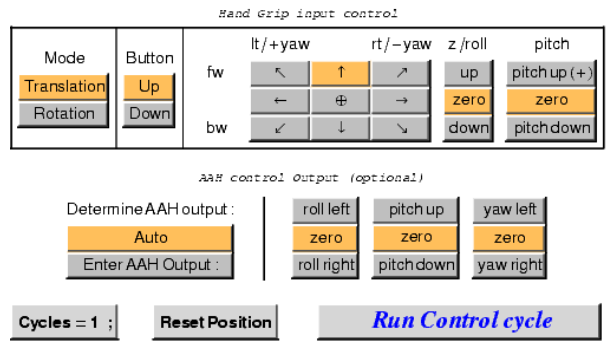}}
\caption{The GUI for the hand controller.}
\label{gui}
\end{center}
\end{figure}
An interface to the hand controller similar to that in \cite{Agerholm&97b} 
is provided in Mathematica (Figure \ref{gui}). It contains buttons for all the 
hand controller states as well as for manual input of the AAH output for 
overriding the simulated AAH in the model.

Pressing one of the buttons sets a global variable that is used to determine 
the parameters passed to the {\tt ControlCycle} function. Additionally, the 
"Cycles=1" button determines how many control cycles should be evaluated 
when the "Run Control Cycle" button is pressed.

\begin{figure}[h]
\begin{center}
\begin{minipage}{6.7cm}
\resizebox{6.7cm}{!}{\includegraphics{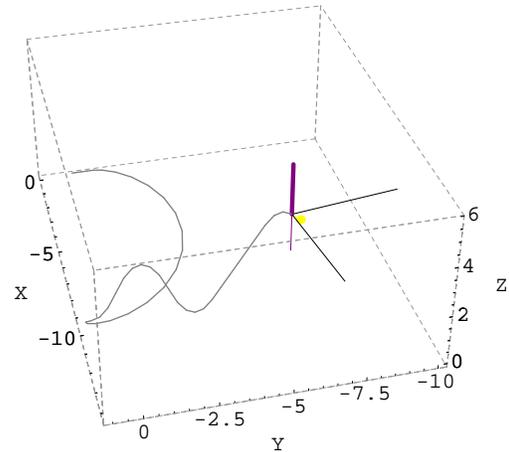}}
\end{minipage}
\caption{A sample trajectory of the SAFER.}
\label{path}
\end{center}
\end{figure}

Pressing "Run Control Cycle" initiates the control cycle and after calculating
the new position prints out a plot of the astronaut's path so far together with
his orientation indicated by the axes of his own coordinate system (Figure \ref{path}).
Additionally, his velocity and angular velocity are shown as vectors. Optionally
a table with the list of the fired thrusters as well as the axes where AAH is 
turned on is printed.

Since all the previous position data is stored, Mathematica can even animate 
this graph so that one can inspect the SAFER moving through space.

\begin{figure}[h]
\begin{center}
\begin{boxedminipage}[h]{3.0938in}
{\scriptsize
\begin{alltt}
ResetSAFERPosition[ ];
(* 1 right *)
Do[SensorControlCycle[SwitchPositions[TRAN,UP], 
    HandGripPosition[ZERO,ZERO,POS,ZERO]],\{1\}];
(* 3 yaw *)
Do[SensorControlCycle[SwitchPositions[ROT,UP], 
    HandGripPosition[ZERO,ZERO,POS,ZERO]],\{3\}];
(* 15 "right" *)
Do[SensorControlCycle[SwitchPositions[TRAN,UP], 
    HandGripPosition[ZERO,ZERO,POS,ZERO]],\{15\}];
(* wait *)
Do[SensorControlCycle[SwitchPositions[TRAN,UP], 
    HandGripPosition[ZERO,ZERO,POS,ZERO]],\{2\}];
(* 3 up *)
Do[SensorControlCycle[SwitchPositions[TRAN,UP], 
    HandGripPosition[POS,ZERO,ZERO,ZERO]],\{3\}];
(* 6 down *)
Do[SensorControlCycle[SwitchPositions[TRAN,UP], 
    HandGripPosition[NEG,ZERO,ZERO,ZERO]],\{6\}];
(* 5 up *)
Do[SensorControlCycle[SwitchPositions[TRAN,UP], 
    HandGripPosition[POS,ZERO,ZERO,ZERO]],\{5\}];
(* nothing, just keep floating in space *)
Do[SensorControlCycle[SwitchPositions[TRAN,UP], 
    HandGripPosition[ZERO,ZERO,ZERO,ZERO]],\{6\}];
(* finally, 2 down *)
Do[SensorControlCycle[SwitchPositions[TRAN,UP], 
    HandGripPosition[NEG,ZERO,ZERO,ZERO]],\{2\}];
\end{alltt}
}
\end{boxedminipage}
\end{center}
\caption{The commands to create the sample trajectory.}
\label{cycle}
\end{figure}

A graphical interface to the simulation like in Figure \ref{gui} is
interesting when testing the system's behavior in general. However,
when adjusting parameters or testing specific cases, it's more
convenient to run the control cycles directly using Mathematica input
commands. Figure \ref{cycle} shows the input to create
Figure~\ref{path}.

In \cite{Agerholm&97} the visualization is
done outside the toolbox using dynamic link modules, which are programmed
specifically for this one application. In Mathematica, changing only the
differential equations suffices to include other influences like gravity, as 
Mathematica chooses the algorithm to solve the equations.\\

However, testing in Mathematica is not restricted to graphical
simulation.  Like in \cite{Agerholm&97}, the output of the thruster
selection logic can be validated by enumerating all possible states of
the Hand controller, or in an extended version enumerating all
possible states of the hand controller and the AAH.  Figure \ref{tests}
shows these functions formulated in Mathematica's VDM-SL notation.
On every possible
state, ControlCycle is applied to calculate the fired thrusters. The
result of this large map comprehension then has to be investigated
manually.

\begin{figure}[h]
\begin{center}
\begin{boxedminipage}[h]{3.0938in}
{\scriptsize
\begin{alltt}
VDMFunction[  ControlCycleTest,
  SwitchPositions \(\times\) HandGripPosition \(\times\) RotCommand ->
    ThrusterSet,
  ControlCycleTest[SwitchPositions[mode_, aah_], rawGrip,
      aahCmd]:=
    SelectedThrusters[HCM`GripCommand[rawGrip, mode], 
      aahCmd, AAH`ActiveAxes[], AAH`IgnoreHcm[]],
  True,
  card[RESULT]\(\leq\) 4 \(\land\) ThrusterConsistency[RESULT]
];

VDMFunction[  BigTest,
  \{\}->(HCM`SwitchPositions \(\times\) HCM`HandGripPosition \(\times\)
        AUXIL`RotCommand -> ThrusterSet),
  BigTest[]:= map[(\{switch, grip, aahLaw\}->
      ControlCycleTest[switch, grip, aahLaw])|
      \{switch\(\in\)switchPositions, grip\(\in\)gripPositions,
      aahLaw\(\in\)allRotCommands \}]
]

VDMFunction[  HugeTest,
  \{\}->(HCM`SwitchPositions \(\times\) HCM`HandGripPosition \(\times\)
        AUXIL`RotCommand -> ThrusterSet),
  HugeTest[]:= map[(\{switch, grip, aahLaw\}->
      ControlCycleTest[switch, grip, aahLaw])|
      \{switch\(\in\){}switchPositions, grip\(\in\){}allGripPositions,
      aahLaw\(\in\){}allRotCommands\}]
];
\end{alltt}
}
\end{boxedminipage}
\end{center}
\caption{The testing functions.}
\label{tests}
\end{figure}

Another important part in the process of verifying software would be 
coverage testing, which is unfortunately not possible in Mathematica.

\section{Enhanced Analysis of the System}
\label{enhanced}

The simulation possibilities described in the last section can 
be exploited for risk and  safety analysis of the system. A very 
simple application is the case when one of the thrusters fails 
due to a mechanical defect or an iced valve. The most important 
questions in this scenario are whether the astronaut will still 
be able to navigate the system, and whether it is possible to 
return before the air or the nitrogen for the thrusters is used 
up.

\begin{figure}[h]
\begin{center}
\resizebox{8cm}{!}{\includegraphics{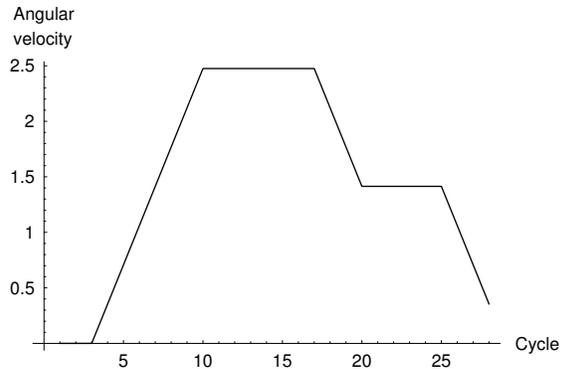}}
\caption{Angular velocity with a broken thruster, AAH initiated in cycle 9.}
\label{AngularVelocityDiagram}
\end{center}
\end{figure}

We investigated the functionality of AAH in the case where one
thruster ({\sf 6-F2}) fails. Figure \ref{AngularVelocityDiagram} shows
the angular velocity of the system, with the hand grip set to forward
acceleration. Just before cycle 4 is initiated, thruster {\sf 6-F2}
breaks, which would be used in this acceleration. This leaves thruster
{\sf 7-F3} applying an additional torque to the system, which results in an
increasing angular velocity. In cycles 9 and 10 the astronaut
initiates AAH, but keeps the forward acceleration (cycles 10 to 17 and
20 to 25). AAH is now only able to compensate the additional torque,
but not to reduce the angular velocity. Only when the forward
acceleration is turned off (cycles 17 to 20 and 25 to 30), AAH shows
effect.

The functionality of AAH could be improved by immediately 
excluding thruster {\sf 7-F3} from the translational commands when 
thruster {\sf 6-F2} fails (and thus allowing thruster {\sf 3-B3} to be used 
by AAH instead of {\sf 6-F2}). 
This would require a slightly modified and more 
complex thruster selection logic, providing a higher level of safety 
for the astronaut. 

\section{Concluding Remarks}
\label{conclusion}
In this article a hybrid model of NASA's SAFER system has been presented 
using the specification language VDM-SL inside the computer algebra system
Mathematica. We demonstrated that the implementation of a VDM-SL package for
Mathematica provides both, VDM-SL's powerful language features, like 
comprehensions, as well as the mathematical power of Mathematica, e.g.  
solving differential equation systems. 
 
The SAFER example shows the validation possibilities of such a combined tool. 
Like in 
\cite{Agerholm&97} the complex discrete model of the control logic can be
validated through testing. This is a cheap technique for raising 
the confidence that the right model has been specified prior to the 
application of more expensive formal proof techniques. 

However, with the right tool, there is no reason why the continuous
models of a hybrid system should be excluded from validation.  Such a
hybrid validation is more suitable for finding unjustified domain
assumptions made in the discrete model.  We strongly propose such
validations, due to the fact that making wrong assumptions is the weak
point of formal verification techniques, possibly leading to correct
proofs of the wrong model.

Furthermore, we demonstrated that the visualization features of Mathematica 
provide a convenient way to communicate a model to a customer. Moreover, in
contrast to \cite{Agerholm&97}, our visualization is a functional graph that
facilitates the communication to control experts as well as to customers
with a technical expertise. 

In the Irish school of VDM, Mathematica has been used to
explore explicit VDM specifications \cite{Reilly95}, but
to our present knowledge not for modeling hybrid systems.  
 
Note that the conclusion of our work is not that Mathematica is the best tool
for validating hybrid system specifications. Our Mathematica approach has its 
disadvantages, too: Our VDM-SL representation is not as readable as the
notation of standard VDM-SL and a typed language would be more suitable for 
specification purposes.
Rather than proposing a certain tool, our work points out the features a 
powerful toolset should provide for validating hybrid systems.

Another future approach would be the integration of a classic formal 
method tool
with a computer algebra system. For example a combination of Mathematica
with the IFAD VDM-SL Toolbox used in \cite{Agerholm&97} would be a possibility.
This could be realized with the lately developed CORBA API of this tool,
that enables access to the toolbox as a CORBA object and thus
calling its VDM-SL interpreter from programs implemented in C or Java. 
Mathematica provides an interface through its MathLink facility.

Summarizing, we feel that our approach of hybrid validation is a
valuable technique for producing systems of higher reliability
and hope that it will stimulate further research in this area.

\subsection*{Acknowledgment}
Many thanks to William Milam from the Ford Motor Company. At the FME'96 
conference, he pointed the first author to the industrial needs of 
analytical methods and tools for hybrid systems. 
Peter Gorm Larsen and Peter Lucas were kind enough to comment on a draft of 
this paper for which we are very thankful.
Finally, the authors would like to thank the four anonymous referees for the 
interesting comments and suggestions.  

\bibliographystyle{plain}
\bibliography{lfm2000}
\end{document}